\def\mode{0} 

\if 1\mode
    \documentclass[preprint,superscriptaddress,floatfix]{revtex4-2}
    \usepackage{lineno}
    \linenumbers
    \usepackage{natbib}
\else
    \documentclass[twocolumn,superscriptaddress,nofootinbib,runinaddress]{revtex4-2}
\fi

\usepackage{silence}
\usepackage{graphicx}
\usepackage{dcolumn}
\usepackage{bm}
\usepackage{caption}
\usepackage{subcaption}
\usepackage{float}
\usepackage{siunitx}
\usepackage{amsmath,amssymb}
\usepackage[
	pdftex,
	colorlinks=true,
  	urlcolor=blue,
  	linkcolor=black,
  	citecolor=black
]{hyperref}
\renewcommand{\vec}[1]{\ensuremath{\bm{#1}}}

\newcommand{\ket}[1]{\ensuremath{\left\vert{#1}\right\rangle}}
\newcommand{\bra}[1]{\ensuremath{\left\langle{#1}\right\vert}}

\usepackage{mathptmx}

\begin{document}

\title{Spin Squeezing with Magnetic Dipoles}
\author{Alec Douglas}
\thanks{These authors contributed equally to this work}
\author{Vassilios Kaxiras}
\thanks{These authors contributed equally to this work}
\author{Lin Su}
\author{Michal Szurek}
\author{Vikram Singh}
\author{Ognjen Markovi\'c}
\author{Markus Greiner}
\if 0\mode
    \thanks{amdouglas@g.harvard.edu, vkaxiras@caltech.edu,\\ greiner@physics.harvard.edu}
\fi
\affiliation{Department of Physics, Harvard University.}

\date{\today}

\begin{abstract}
Entanglement can improve the measurement precision of quantum sensors beyond the shot noise limit. Neutral atoms, the basis of some of the most precise and accurate optical clocks and interferometers, do not naturally exhibit the all-to-all interactions traditionally used to generate such entangled states. Instead, we take advantage of the magnetic dipole-dipole interaction native to most neutral atoms to realize spin-squeezed states. We achieve 7.1 dB of metrologically useful squeezing using the finite-range spin exchange interactions in an erbium quantum gas microscope. d demonstrate that introducing atomic motion protects the spin sector coherence at low fillings, significantly improving the achievable spin squeezing in a 2D dipolar system. This work's protocol can be implemented with most neutral atoms, opening the door to quantum-enhanced metrology in other itinerant dipolar systems, such as molecules or optical lattice clocks, and serves as a novel method for studying itinerant quantum magnetism with long-range interactions.
\end{abstract}

\maketitle

\begin{figure*}[htp]
    \includegraphics[width=0.98\textwidth]{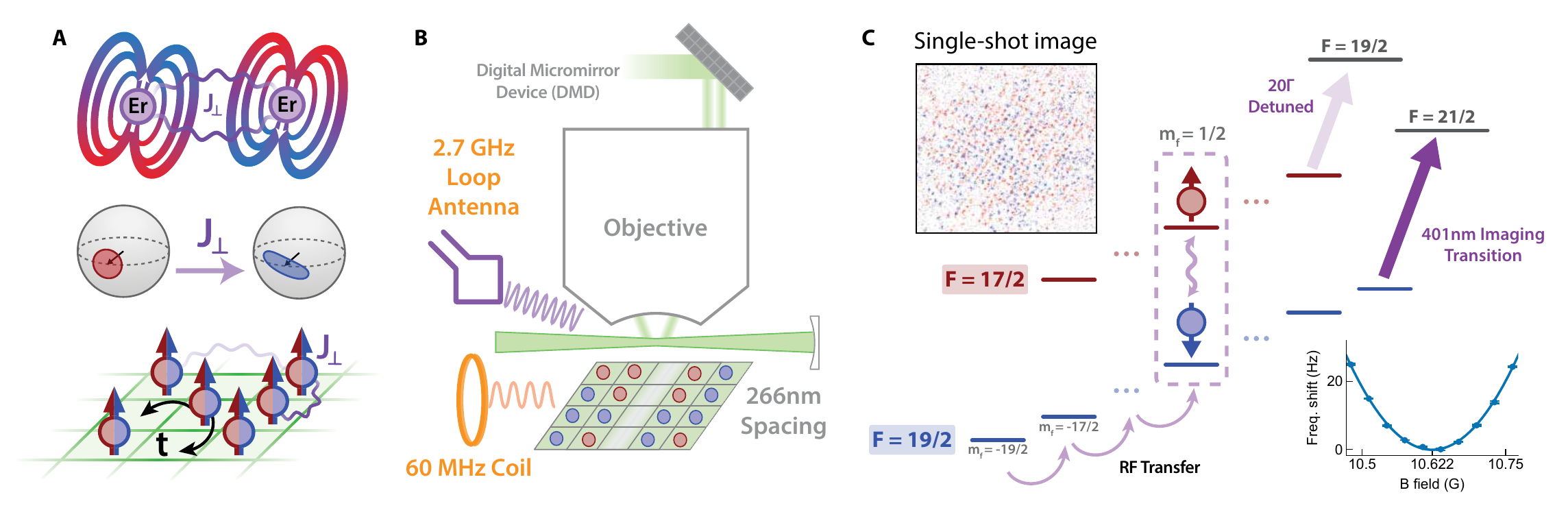}
    \caption{
    \textbf{Spin squeezing using magnetic dipole interactions between erbium atoms in a quantum gas microscope}
    \textbf{a}, Our experiment utilizes the intrinsic magnetic dipole moment of $^{167}$Er to generate entanglement. The interaction shears an initial coherent state, squeezing it along one axis. We load a 2D layer of atoms in our optical lattice, where the finite-range spin-exchange interaction $J_{\perp}$ is used to generate entanglement. Optionally, the atoms can tunnel to adjacent sites with tunable rate $t$. 
    \textbf{b}, Atoms in a 266 nm spacing lattice are loaded into tunable shapes by a Digital Micromirror Device (DMD). The atom's spin states are manipulated with a 2.7 GHz loop antenna and an in-vacuum coil at 60.8 MHz. 
    \textbf{c}, The atoms are prepared in the $m_F=-19/2$ state, and are transferred to $m_F=1/2$ (the qubit manifold) through successive Landau-Zener sweeps. The magnetic field is set such that the qubit frequency is first-order insensitive to magnetic field fluctuations, as shown in the bottom right of the subfigure. We selectively image the two spin manifolds by a differential 20$\Gamma$ detuning of the imaging transition. The two spin states are shown as red and blue dots in a single-shot image on the left.}
    \label{fig:mainexp}
\end{figure*}

\section{Introduction}

Entangled states allow quantum sensors to overcome the Standard Quantum Limit (SQL), and have enhanced optical clocks \cite{pedrozo-penafiel_entanglement_2020,robinson_direct_2024}, atom interferometers \cite{greve_entanglement-enhanced_2022}, and gravitational wave detectors \cite{jia_squeezing_2024}. One prominent method to generate large-scale entanglement is spin squeezing, where uncertainty is redistributed between quadratures of the collective spin of an atomic, solid state, or molecular ensemble, improving metrological precision \cite{kitagawa_squeezed_1993,ma_quantum_2011,wineland_spin_1992, degen_quantum_2017}. In cold atom systems, spin squeezing generation has been extensively studied in the context of all-to-all interactions or measurements, such as in single mode cavities \cite{appel_mesoscopic_2009,leroux_implementation_2010,schleier-smith_states_2010,hosten_measurement_2016, cox_deterministic_2016} and single mode Bose Einstein Condensates \cite{esteve_squeezing_2008, riedel_atom-chip-based_2010,gross_nonlinear_2010,lucke_twin_2011,hamley_spin-nematic_2012}. Recently, it was predicted that scalable spin squeezing can be achieved with finite range interactions \cite{perlin_spin_2020,gil_spin_2014,foss-feig_entanglement_2016}. This is remarkable from a fundamental perspective, as the nonequilibrium squeezing dynamics may be linked to equilibrium symmetry-broken phases with long-range order \cite{block_scalable_2024}. Additionally, squeezing from finite range interactions is very attractive for metrological applications, as such interactions are inherently present in many cold atom, molecule, and solid-state systems \cite{chomaz_dipolar_2022,cornish_quantum_2024,browaeys_many-body_2020,wolfowicz_quantum_2021}. Recently, squeezing was realized through induced Rydberg interactions, of dipolar \cite{bornet_scalable_2023} or van der Waals \cite{eckner_realizing_2023,hines_spin_2023} nature, and in trapped-ion platforms with tunable power-law interactions \cite{franke_quantum-enhanced_2023}. 

In this work, we demonstrate spin squeezing through magnetic dipole-dipole interactions. Such interactions are available natively in most atomic species. The spin squeezing is carried out in an erbium quantum gas microscope, in which we prepare $\approx 10^{3}$ itinerant XY interacting fermions in an optical lattice. First, we describe novel methods for generating and imaging magnetically insensitive dipolar interacting spin states in fermionic $^{167}$Er, inspired by Zou et al. \cite{zou_magnetic_2020}. While benchmarking the platform, we observe indications of the ferromagnetic ordering expected of the many-body ground state at high filling, predicted to lead to an extensive squeezed state \cite{foss-feig_entanglement_2016,perlin_spin_2020,block_scalable_2024,comparin_scalable_2022}. Next, we generate and directly measure metrologically useful squeezing up to $\qty{7.1(1.0)}{dB}$ with no corrections for experimental imperfections.

Finally, we introduce a new degree of freedom, itinerancy, which has recently been proposed to enhance squeezing in finite-range interacting systems \cite{bilitewski_dynamical_2021,wellnitz_spin_2024}. Although finite lattice fillings impede the formation of an ordered ground state \cite{kwasigroch_synchronization_2017} and thus a squeezed state, tunneling was recently numerically shown to help overcome this effect \cite{wellnitz_spin_2024}. However, it remains an open question whether this enhancement survives at finite temperature and finite tunneling, where the partial delocalization of defects may impede local ordering. Additionally, the study of the interplay between tunneling and power law interactions has been of substantial recent interest due to advances in ultracold molecule experiments  \cite{li_tunable_2023,lee_spin_2023,carroll_observation_2024}. Remarkably, we observe that in dilute systems where there is no metrological gain when the spins are static, the introduction of itinerancy protects the long range order and leads to metrologically useful squeezing of \qty{5.2(0.6)}{dB}.

\section{Experimental Platform and Squeezing Theory}
We generate spin squeezing using a magnetic dipolar spin exchange, or XY model, between two hyperfine states (Fig. \ref{fig:mainexp}a). For sufficiently high fillings, and for sufficiently slowly decaying power-law interactions, the system chooses a global phase, creating a U(1) symmetry-broken ferromagnet \cite{kwasigroch_synchronization_2017,chen_continuous_2023}. This ferromagnet is suspected to protect the permutational symmetry of the ensemble, leading to effective dynamics restricted to the Dicke manifold \cite{block_scalable_2024}. For short times, the dipolar XY model has recently been shown numerically to produce similar macroscopic dynamics to the all-to-all one axis twisting (OAT) model \cite{perlin_spin_2020,comparin_multipartite_2022}: $H_{\mathrm{OAT}}=\chi \hat{S}_z^2=\frac{\chi}{4}\sum_{i,j}\hat{\sigma}_{z,i}\hat{\sigma}_{z,j}$, where $\hat{S}$ is the collective spin of $N$ atoms restricted to a pseudo spin-1/2 qubit basis. The emergence of a $\hat{S}_z^2$ term can be seen by rewriting the XY Hamiltonian as $\hat{S}_x^2+\hat{S}_y^2=\hat{S}^2-\hat{S}_z^2$, and noting ferromagnetic states are eigenstates of the total spin $\hat{S}$, allowing us to ignore  $\hat{S}^2$ in dynamics \cite{perlin_spin_2020}. Qualitatively, the dynamics of OAT rotate the total spin vector $\hat{S}$ around the $z$ axis at a rate proportional to $\langle \hat{S}_z\rangle$. When acting on a state polarized along the equator of the generalized Bloch sphere, this shears the quantum projection noise (QPN) in different directions, resulting in a squeezed state (Fig.\ref{fig:mainexp}a). The squeezing typically increases until the optimal squeezing time, after which over-squeezing impedes metrological use. 

We implement the dipolar XY model in an erbium quantum gas microscope (QGM) \cite{su_dipolar_2023}. Although QGMs are typically designed as quantum simulators, they have many advantages when used for metrology. Most notably, QGMs can natively scale to large atom numbers, support site-resolved imaging, and project arbitrary spatial potentials on atomsWe use tightly spaced \qty{266}{nm} optical lattices in the $x$ and $y$ directions to maximize the dipolar interaction and trap our $^{167}$Er atoms in one vertical plane of a \qty{532}{nm} spacing $z$ lattice. $^{167}$Er has a rich hyperfine structure, with 104 total $F,m_F$ levels. We perform all of our experiments in a single two-level qubit manifold: $\ket{\downarrow}=\ket{F=19/2, m_F=1/2}$, and $\ket{\uparrow}=\ket{17/2,1/2}$. We choose hyperfine states to maximize the product of the qubit's coherence time with its interaction strength. Rather than choosing the fastest interacting states, we use a transition that is first-order magnetically insensitive at \qty{10.6}{Gauss} which allows us to effectively eliminate magnetic field noise as a source of decoherence (see Methods and Fig. \ref{fig:mainexp}c). Additionally, our chosen states have similar or lower interactions than less magnetic atoms, allowing for direct extension of our work to other neutral atoms (see Methods).
\begin{figure*}[ht]
    \includegraphics[width=\textwidth]{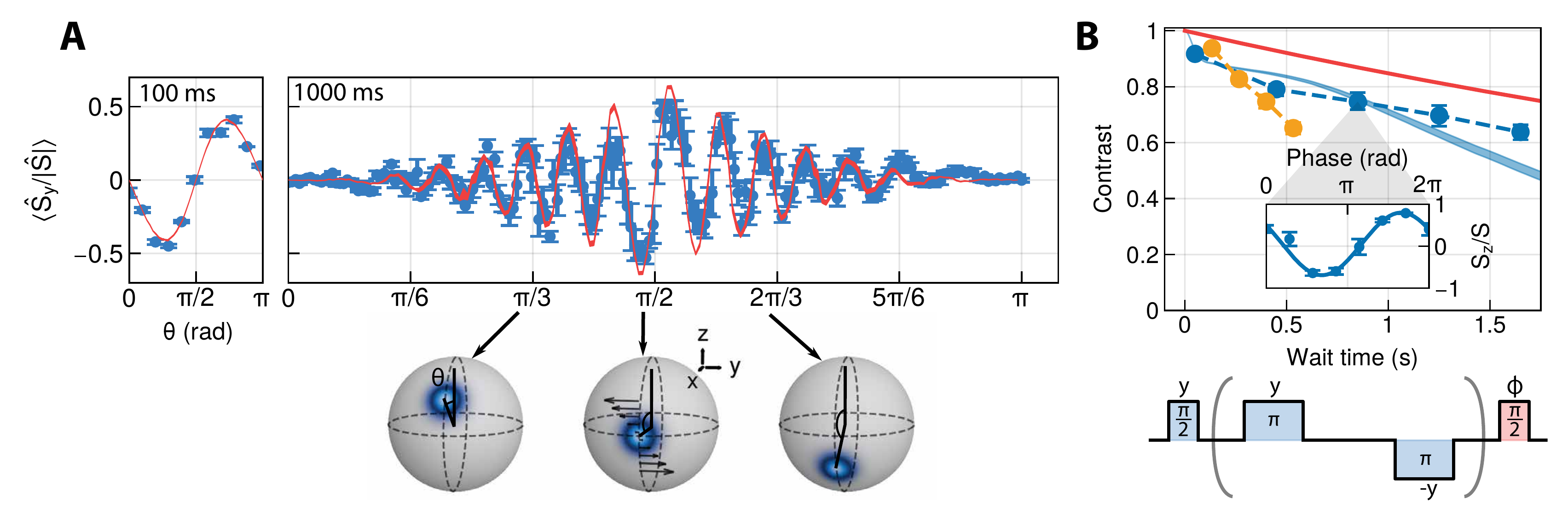} 
    \caption{
    \textbf{Characterization of the spin-spin interaction.}
    \textbf{a}, We observe the XY interaction by the shearing it induces in the generalized Bloch sphere of 600 (1100) atoms at 100 ms (1 s). Each blue point corresponds to a different initial state, parameterized by $\theta$, as shown in the Bloch spheres below the data.
    \textbf{b}, Ramsey contrast decay over time in a cloud of 1000 atoms with peak fillings of $\sim 85$\% (blue), 85 atoms with center filling $\sim 20$\% (orange), and the fitted interaction-free $T_2$ coherence (red). The orange and blue data, taken without decoupling the interaction, lie below the red curve, showing that our experimental coherence is interaction-limited. Each point is the fitted amplitude of a Ramsey fringe (see inset), and the shaded curve is DTWA.
    }
    \label{fig:mean}
\end{figure*}
Our quantum simulator realizes the effective 2D XY dipolar Hubbard model with itinerant fermions (Fig. \ref{fig:mainexp}a). The on-site interaction between different spin species is much larger than all other scales (Materials and Methods), therefore, we truncate the Hamiltonian to one particle per site. The effective Hamiltonian is then:
\begin{equation}
\label{eq:Hamiltonian}
H=-t\sum_{\langle i,j \rangle,\sigma}(\hat{c}_{i,\sigma}^\dagger \hat{c}_{j,\sigma}+\mathrm{h.c.})+J_{\perp}\sum_{i\neq j}\frac{a^3}{r_{i,j}^3} \hat{S}^{+}_i \hat{S}^{-}_j
\end{equation}
Here, $t$ describes the tunneling of hard-core fermions along nearest neighbor bonds $\langle i,j \rangle$ and $J_{\perp}$ the characteristic dipolar exchange energy between adjacent sites, with a fixed value of \qty{1.09}{Hz} throughout this work. $r_{i,j}$ is the distance between sites $i$ and $j$, and $a=\qty{266}{nm}$ is the characteristic lattice spacing. Finally, $\hat{c}^{\dagger}_{i,\sigma},\hat{c}_{i,\sigma}$ are the fermionic creation/annihilation operators with spin index $\sigma$, and $\hat{S}_i^{+}=\hat{c}^\dagger_{i,\uparrow} \hat{c}_{i,\downarrow}$ ($\hat{S}_i^{-}=\hat{S}_i^{+\dagger}$) the spin raising (lowering) operators. The quantization axis is kept perpendicular to the plane, which ensures isotropic dipolar interactions. The states we choose have minimal differential Stark shifts due to lattice light at our chosen polarization, leading to mHz level disorder in the spin sector (see Methods). We also keep $t$ either much larger than the site-to-site disorder of \qty{3}{Hz} RMS or set tunneling much smaller than any other energy scale ($<\qty{.3}{Hz}$), allowing us to ignore disorder in the kinetic sector.

Each experiment begins by adiabatically loading a spin-polarized band insulator from a degenerate Fermi gas of the stretched hyperfine state $\ket{g} =\ket{19/2, -19/2}$. The insulator is shaped into various spatial configurations by tuning the local chemical potential with a digital micro-mirror device (DMD) during the loading (Fig.\ref{fig:mainexp}b). We then perform a series of Landau-Zener sweeps to transfer the spin state from $\ket{g}$ to $\ket{\downarrow}$, using an in-vacuum coil at 60.8 MHz (Fig. \ref{fig:mainexp}c and Methods). Next, we transfer the atoms from $\ket{\downarrow}$ to $\ket{\uparrow}$ using 2.7 GHz microwave pulses. We ensure spin purity by removing atoms in all states in the $F=19/2$ manifold with a $\qty{583}{nm}$ resonant beam, retaining only the atoms transferred to $\ket{\uparrow}$. At the end of this process, we are left with peak fillings of about 90\% and atom numbers near 1000, dominated by initial loading (Fig. \ref{fig:mainexp}c).

The atoms are imaged with single-site resolution by transferring them to an accordion lattice \cite{su_fast_2024}. We image both spin states with sequential images, transferring states between images, as imaging light is mostly resonant with the $\ket{\downarrow}$ state (Fig. \ref{fig:mainexp}c). The fidelity is further enhanced by shelving atoms in states that couple primarily to opposite polarizations of light. We measure a \qty{99.5}{\%} fidelity to distinguish between 0 and 1 atom, and a loss of the second state due to off-resonant scattering from the first image of less than \qty{.25}{\%} (see Methods). 
\begin{figure*}[htp!]
    \includegraphics[width=\textwidth]{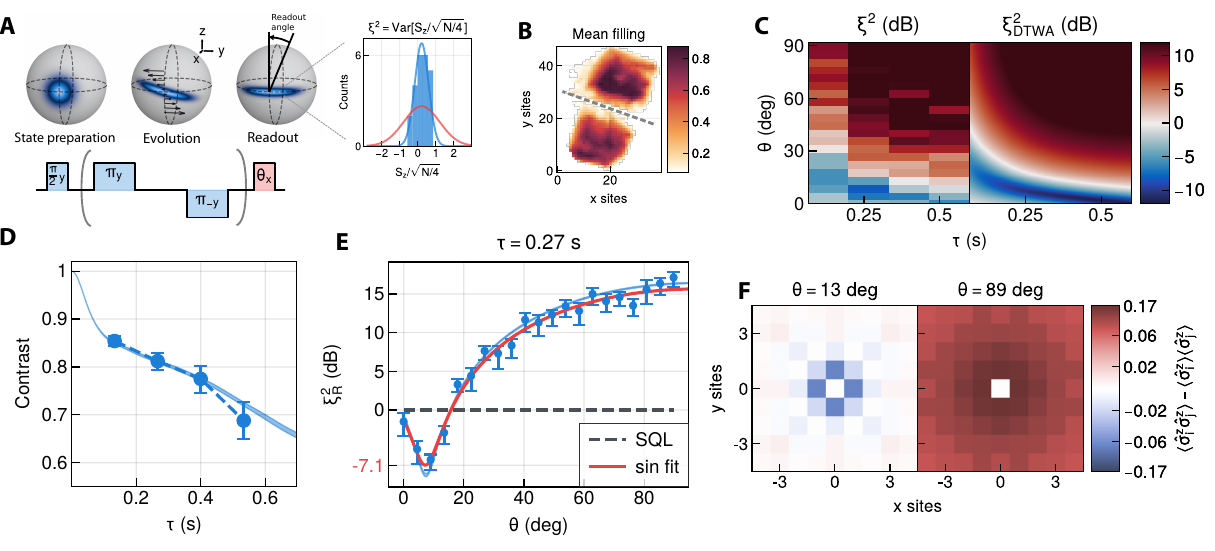}
    \caption{
    \textbf{Spin squeezing from finite range magnetic XY interactions}
    \label{fig:squeeze}
    \textbf{A}, The interaction, shears the collective spin along the black arrows on the Bloch sphere. We estimate the noise squeezing $\xi^2$ by sampling $S_z$. A histogram of $S_z/\sqrt{N/4}$ from the most squeezed point in e, with a variance of $\xi^2=$-8.2$\pm$1.0~dB below the SQL, is shown. A Gaussian with the measured (SQL) variance is shown in blue (red).
    \textbf{B}, We perform differential measurements of $\xi^2$ with two spatially separated 260 atom clouds to reject common-mode noise.
    \textbf{C}, The noise squeezing is sampled over evolution time and readout angle. The left square is experimental data, the right is a numerical simulation (DTWA) based on the filling in b.
    \textbf{D}, We see excellent agreement in the Ramsey contrast with DTWA (shaded curve) at all sampled times. 
    \textbf{E}, $\xi^2_R$ after $\tau=0.27$~s of evolution. The optimal squeezing measured was -6.4$\pm$1.0~dB, and fitted to be -7.1$\pm$1.0~dB.
    \textbf{F}, We visualize the squeezing and antisqueezing by computing site-to-site two-particle spin correlations radially averaged over the clouds at a squeezed and an antisqueezed readout angle.
    }
\end{figure*}
\section{Spin Control and Mean-Field Interaction}

The loading procedure initializes a high-filling array of spin polarized, static fermions in a short spacing lattice. We then use the magnetic dipolar interaction between states in the atomic hyperfine manifold to realize squeezing dynamics. Such interactions were previously observed as a mean field shift in $^{87}$Rb \cite{zou_magnetic_2020}. To study the interaction and show that it can generate coherent many-body dynamics we prepare the atoms in an initial coherent superposition with $\langle \hat{S}_z\rangle=\cos\theta$, for $\theta\in[0, \pi]$. We observe the shearing effect of the effective OAT Hamiltonian by measuring $\langle \hat{S}_y\rangle$ to track the precession of the mean spin vector. This is done by exchanging $\hat{S}_y$ and $\hat{S}_z$ with a $\pi/2$ pulse before readout. Fig. \ref{fig:mean}a shows the onset of the interaction at short time, and at a time far beyond the optimal squeezing time to characterize the coherence of the state evolution. We compare our results with simulations with a parameter-free Discrete Truncated Wigner Approximation (DTWA; see Methods), using the theoretically predicted exchange rate of \qty{1.09}{Hz}, and observe agreement at all times sampled in this work.

A prerequisite for useful spin squeezing is a high degree of coherence, or equivalently a ferromagentic ordering of the constituent spins. In particular, with a dipolar XY interaction, squeezing is achievable with states prepared near the ferromagnetic ground state \cite{block_scalable_2024,bornet_scalable_2023}. Thus, by probing the dynamics of its coherence, we validate that our system can spin squeeze via an XY model. For no interaction, noise in the classical control field will limit the lifetime of a spin-polarized state, representing the upper bound of our coherence. We implement an interaction decoupling WAHUHA + spin echo pulse sequence \cite{choi_robust_2020} and report a $T_2$ of $\qty{6.0(2)}{s}$ (see Methods, Fig. \ref{fig:latticeloss}a). Next, we exclusively decouple frequency fluctuations using spin echo pulses and compute the Ramsey contrast $C\equiv \frac{\langle|\hat{S}|\rangle}{N/2}$, which depends on both the presence of the interaction and density (Fig. \ref{fig:mean}b). We expect that for stochastic and dilute fillings of the optical lattice, the interaction inhomogeneity will cause the coherence to rapidly decay. At higher densities, the interaction is sufficiently homogeneous to force the atoms to all precess together resulting in long-lived coherence \cite{kwasigroch_synchronization_2017}. For a dataset with a lower atom number ($N=85$) and lower filling ($\sim 20\%$) (orange curve), we observe rapid decay out to \qty{.5}{s}, about one interaction time. For a cloud of $N=1000$ atoms (blue curve), two regimes are present in the decay of $C$: first, for times less than \qty{.2}{s}, a rapid relaxation that we attribute to local equilibration from the interaction, dominated by finite filling and edge effects; and second, a plateau in the coherence, where the system maintains ferromagnetic order over time scales of $1/J_{\perp}$, with the contrast decaying principally due to the generation of antisqueezing as predicted in DTWA. The rapid relaxation comes from the proliferation of finite momentum spin waves during equilibration. However, recent work has shown that these decouple from the dynamics of the zero momentum spin waves that we expect to generate squeezing and preserve coherence \cite{roscilde_entangling_2023}. 

\section{Spin Squeezing}

The observed dipolar shearing spin dynamics together with the preservation of coherence under the dipolar interaction create an excellent starting point for metrological spin squeezing. We now directly measure the squeezing in our system by sampling the distribution of the collective spin of the atoms from QGM snapshots. A set of $N$ uncorrelated spins $\hat{S}=\sum_{i=1}^N\hat{s}_i$ when sampled along an axis $\vec{n}_\perp$ perpendicular to the mean spin $\langle \hat{S}\rangle$, will have fluctuations due to QPN. The degree of squeezing is typically quantified with the squeezing parameter $\xi^2$, which we refer to as noise squeezing:
\begin{equation}
    \xi^2\equiv\frac{4\:\text{min}_{\vec{n}_\perp}\text{Var}[\hat{S}\cdot\vec{n}_{\perp}]}{N}
\end{equation}
where the minimum is over all axes perpendicular to the mean spin direction \cite{kitagawa_squeezed_1993}. A value of $\xi^2<1$ indicates that fluctuations are squeezed below the SQL, as shown in the inset of Fig. \ref{fig:squeeze}a. While $\xi^2<1$ implies entanglement for spin-polarized states, the phase precision of an atomic sensor is only improved by squeezing if the reduction in the spin fluctuations is greater than the reduction in the contrast $C$. Thus metrologically relevant squeezing is typically quantified through the Wineland squeezing parameter $\xi^2_R\equiv\xi^2/C^2$ \cite{wineland_spin_1992}.

To generate and measure a squeezed state we begin by preparing a coherent state on the equator of the generalized Bloch sphere with a $\pi/2$ pulse. After some evolution time $\tau$ we apply a readout pulse of angular length $\theta$ and phase 90 degrees offset from the initialization pulse. During evolution, spin-echo pulses are applied every 66 ms to decouple phase errors from our microwave drive and from lattice Stark shifts. At the end of the sequence, we measure in the $z$ basis via population number measurements, which effectively reads out $\langle\cos(\theta)\hat{S}_z+\sin(\theta)\hat{S}_y \rangle$, the variance of which is used to estimate $\xi^2$. This sequence is pictured in Fig. \ref{fig:squeeze}a.

To account for drifts in the final pulse phase that lead to non-zero shot-dependent $\langle \hat{S}_z\rangle$ and other sources of classical noise, we measure a differential observable of two spatially separated clouds subjected to the same drive fields as depicted in Fig. \ref{fig:squeeze}b (see Methods). The clouds remain slightly coupled, leading to the optimal squeezing angle for the differential observable to decrease over time, even past zero (see Methods).

To extract the Wineland parameter, we measure the noise squeezing $\xi^2$ at 4 values of the evolution time $\tau$ and 21 values of the readout angle $\theta$ in two clouds with 260 atoms each and peak fillings of \qty{80}{\%} in Fig. \ref{fig:squeeze}c, and we measure the Ramsey contrast $C$ at the same time points, in Fig. \ref{fig:squeeze}d. Combining both measurements at $\tau=0.27$ s, we plot $\xi^2_R$ as a function of $\theta$ and extract its minimum by fitting the data to a $\sin$ function with an offset in Fig. \ref{fig:squeeze}e. In this manner, we measure a minimum Wineland squeezing of $\xi^2_R=\qty{-6.4(1.0)}{dB}$, and fit a minimum of $\xi^2_R=\qty{-7.1(1.0)}{dB}$, in good agreement with our parameter-free theory, represented by the shaded curve on the same plot (see Methods). In another experiment with two clouds, of $N \approx 500$ each, we measure roughly $\xi^2_R=\qty{-4}{dB}$. This decrease in squeezing is explained by microwave control errors which introduce variance that scales linearly with atom number relative to the SQL. In addition, we notice that the measured optimal squeezing angle as a function of time in Fig.\ref{fig:squeeze}e agrees with DTWA, strongly implying that the system dynamics remain coherent and generate substantially more squeezing that could be measured improved microwave control.

The site-to-site correlations of these highly squeezed states elucidate to what degree our system's magnetic dipole-dipole interactions mimic an effective OAT model. We exploit our QGM's single site and spin-resolved imaging to directly measure two-particle spin correlations $g^{2}_{i,j} \equiv \langle \hat{\sigma}^z_{i}\hat{\sigma}^z_{j}\rangle-\langle \hat{\sigma}^z_i \rangle \langle\hat{\sigma}^z_j\rangle$ at a squeezed and an antisqueezed point in Fig. \ref{fig:squeeze}f. Averaging correlations over the atomic cloud, we observe correlations that decay over the region of interest, directly related to the limited propagation rate of the finite-range $1/r^3$ interaction. This precludes the effective model from being all-to-all for the early times sampled, despite the mean observables replicating OAT.

We reconcile these pictures by considering the local interpretation in which each spin's twisting rate depends on its neighboring spins, allowing local correlations to build up. At finite filling, this introduces a variance in the precession rate based on inevitable microscopic details that add to $\xi^2$, limiting squeezing. We assume that there are no spin-aligning terms at short times protecting global uniformity, unlike some recent theoretical work \cite{perlin_spin_2020}. Despite this, we prepare a system (Fig. \ref{fig:squeeze}c) where the variance in precession rates is negligible, and twisting rates still mimic global OAT dynamics, due to an effective decoupling of finite momentum spin waves from the dynamics of global observables \cite{roscilde_entangling_2023}.

\section{Spin Squeezing With Itinerance}

\begin{figure}[htp!]
    \includegraphics[width=0.46\textwidth]{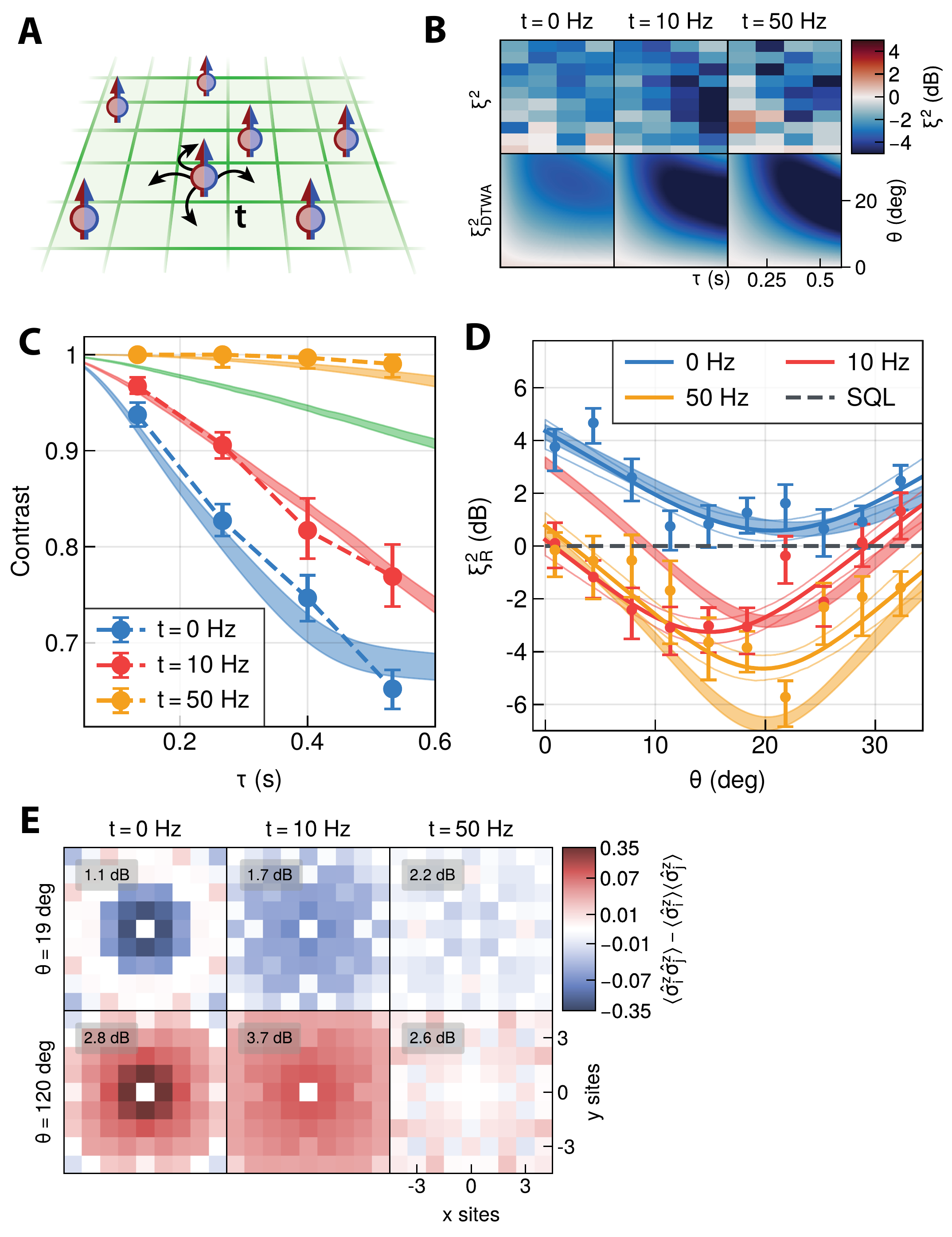}
    \caption{
    \textbf{Itinerancy at low lattice fillings}
    \textbf{a}, Our system supports tunable nearest-neighbor tunneling with rate $t$. We focus on relatively low fillings (20-30\%) to ensure the hardcore fermions do not block each other's movement.
    \textbf{b}, Noise squeezing as a function of evolution time $\tau$, tunneling $t$, and readout angle $\theta$. Tunneling improves noise squeezing in both theory and experiment. Each square shows the same 2D sweep of $\tau$ and $\theta$. Experimental data is in the top row, compared to DTWA in the bottom row. The effects of tunneling are incorporated into DTWA with stochastic hopping for 10 Hz, and One Axis Twisting (OAT) dynamics for 50 Hz.
    \textbf{c}, Ramsey contrast decay as a function of $\tau$ and $t$. At high tunneling (yellow), the contrast remains above \qty{98}{\%} as the system exhibits effective OAT dynamics. In this regime, stochastic tunneling (green curve) is not an accurate model. At intermediate tunneling (red), it is still significantly improved from the no-tunneling case (blue), and is captured by stochastic hopping in DTWA (red curve). The data points are connected with dashed lines to guide the eye.
    \textbf{d}, Wineland squeezing parameter after an evolution time of 0.53 s as a function of tunneling. The squeezing is significantly improved for both $t=10$ and $t=50$ Hz. The data are fitted to quadratic functions.
    \textbf{e}, Radially averaged correlations $g^2_{i,j}$ for $\tau=200$ ms as a function of tunneling. While they fall off rapidly beyond nearest-neighbor at $t=0$, the correlations spread across the lattice for larger $t$. The noise squeezing (antisqueezing) for the whole system at each tunneling is inset in dB for the top (bottom) row.
    }
    \label{fig:tunneling}
\end{figure}

Having successfully generated spin-squeezing at high fillings, we now investigate how itinerancy may enhance the viability of lower-density finite-range interacting platforms. Realizing high filling can be challenging for many up-and-coming molecular platforms with similar dipolar interactions, necessitating the exploration of low-filling regimes. Kinetic degrees of freedom can alleviate interaction inhomogeneity and local disorder in dilute systems. On the other hand, itinerancy may delocalize detrimental defects, introducing time-varying local noise that may disrupt squeezing, as has been observed in recent work with nearest-neighbor interactions \cite{lee_observation_2024}. However, provided the noise spectrum does not couple to the dynamics, tunneling promises to mitigate the limiting factors of finite-range interactions by spatially averaging out imperfections and improving coherence \cite{wellnitz_spin_2024,bilitewski_dynamical_2021}.  

Here we explore the interplay between motional degrees of freedom and the spin squeezing and contrast decay generated by dipolar XY interactions. By tuning the depth of our optical lattice, we probe three regimes of itinerancy where the nearest-neighbor tunneling parameter $t$ is \qty{0}{Hz}, \qty{10}{Hz}, and \qty{50}{Hz}, with peak fillings below 35\%. We measure contrast and $\xi^2$ after 4 evolution times ranging from 133 to 533 ms, and $\theta$ sampled from 0 to 30 degrees. We observe a significant improvement in $\xi^2$, most notably in the later time points, a result reinforced by our simulations (Fig. \ref{fig:tunneling}b). The effects on the Ramsey contrast are even more prominent, as we see the contrast is improved already in the $t=$ \qty{10}{Hz} case but vastly so for $t=$ \qty{50}{Hz} (Fig. \ref{fig:tunneling}c), where it remains above 98\% at all sampled evolution times.

For a fixed evolution time of 533 ms, we compute the Wineland parameter $\xi^2_R$ as a function of readout angle and fit a quadratic to extract its minimum (Fig. \ref{fig:tunneling}d).  At this low filling without tunneling, the Ramsey contrast has sufficiently decayed such that $\xi^2_R$ is never below the SQL. Adding tunneling reduces $\xi^2_R$ by $\qty{3.8(6)}{dB}$ for $t=$ \qty{10}{Hz}, and as much as $\qty{5.2(6)}{dB}$ for $t=$ \qty{50}{Hz}, shifting the curve to well within the regime of useful squeezing.

The spatial distribution of the entanglement across the cloud is measured by the two-particle spin correlations $g^2_{i,j}$ (Fig. \ref{fig:tunneling}e). The spin correlations fall off rapidly for $t=$ \qty{0}{Hz}. At $t=$ \qty{10}{Hz} we see significant delocalization, although the strongest correlations are still nearest-neighbor. At $t=$ \qty{50}{Hz} we observe substantially lower correlations in the window. However, the total squeezing and antisqueezing across the clouds are similar for each value of the tunneling. Therefore the correlations extend much further for $t=$ \qty{50}{Hz}, and indeed we do not observe substantial decay as a function of distance across the averaging window; thus tunneling redistributes the correlations from being appreciable only between neighboring spins to spanning the entire cloud, as would be expected for all-to-all OAT.

We attempt to capture the physics of these effects by modifying our DTWA simulations. For the intermediate tunneling regime ($t=10$ Hz), we incorporate classical diffusive stochastic hopping of atoms localized to individual sites (see Methods). With just this modification, we qualitatively predict both the observed effects of tunneling on the Ramsey contrast and the effects on the noise squeezing $\xi^2$. We understand the improvement in $\xi^2$ as the reduction of the effective variance in the local collective interaction strength through rapid averaging. For the high tunneling regime ($t=50$ Hz), we experimentally observe significantly higher contrast than is predicted by stochastic tunneling-enabled DTWA, suggesting that the diffusive interpretation of the many-body dynamics breaks down. This may be caused by a lower effective motional temperature not represented in our simulation (see Methods). Instead, we compare our high-tunneling data to an all-to-all OAT model with shearing strength equal to the average interaction across all pairs of atoms in the experimentally observed cloud. We simulated this OAT model with DTWA and obtained good agreement on the contrast and squeezing. 

The realization of all-to-all interactions through itinerance is quite surprising. Contact interactions in Bose-Einstein Condensates realize an infinite range model in a single motional mode. However, for itinerant hard-core particles, as occurs for fermions, or bosons in the Mott limit, many motional modes are occupied and therefore can introduce time-varying disorder, disrupting squeezing. The magnetic dipole-dipole interaction, being sufficiently long-range, is seemingly robust to these defects, leading to all-to-all couplings, while contact interactions are not \cite{lee_observation_2024}. Realizing and further exploring a fully all-to-all interaction through itinerance has several advantages. Since permutational symmetry is preserved, the sample remains spin-polarized, preventing collisional decay. In addition, the complete averaging means that the itinerant squeezing is reversible by changing the sign of the interaction, allowing for Heisenberg scaling protocols, despite the randomization of the atomic positions \cite{colombo_time-reversal-based_2022}.

\section{Outlook}

In this work, we have demonstrated that the native magnetic dipole-dipole interactions of neutral atoms, despite their finite-range nature and susceptibility to microscopic details, are excellent for generating strong spin squeezing with applications to quantum metrology. Our mean field observations indicate that the observed squeezing should continue to scale to thousands of atoms given realistic improved phase and amplitude control of the microwave field. We have also investigated the effects of itinerancy, an open question in the context of quantum-enhanced metrology, and found it can be harnessed to improve both spin coherence and squeezing, even reaching the all-to-all OAT limit for dipolar systems. Furthermore, our work requires only itinerant dipoles, suggesting a path to quantum-enhanced metrology in ultracold molecular platforms.

The realized interaction and subsequent squeezing are generalizable to most neutral atoms, and scale with system size. A spin-exchange interaction on a magnetically insensitive transition is present in any atomic state with non-zero electronic and nuclear spins (see Methods). Such an interaction has already been observed through its effect on the microwave clock transition frequency of $^{87}$Rb \cite{zou_magnetic_2020}. Importantly, this is also applicable to optical clocks, as the $^{3}$P$_2$ metastable states of $^{87}$Sr and $^{173}$Yb host magnetically insensitive qubits with twice the interaction strength used in this work (see Methods) \cite{ludlow_optical_2015}. Furthermore, squeezing in a 2D system with $1/r^3$ interactions (such as in this work) is believed to scale with atom number; for as few as $4,000$ atoms, well beyond -20~dB of squeezing is predicted \cite{perlin_spin_2020}, presenting a substantial potential improvement in clock precision.

The platform presented here has broad utility in quantum simulation and metrology. With tunable itinerancy, site-resolved imaging, and high-filling fractions, we can explore the dynamics and phases of itinerant XXZ interacting fermions such as the $t$-$J$ model. Further comprehensive investigation regarding the effect of kinetic temperature on squeezing and ordering may inform strategies for adding itinerancy to systems for metrological improvement. For example, there are untapped regimes where tunneling energies are comparable to the single particle dipolar interaction, achievable by choosing states in $^{167}$Er with up to \qty{4}{Hz} (\qty{9}{Hz}) interaction strength with (without) a magic B field, that should generate entanglement between the kinetic and spin sectors.

Finally, we show that for dipolar interactions, the addition of large tunnelings interpolates between effective $1/r^3$ and all-to-all interactions, enabling the study of spin models with different interaction forms. This work's entanglement generation scheme can be combined with erbium's optical clock transitions that possess attractive properties, or extended to higher spin models \cite{trifa_scalable_2024}. Given the presence of transitions at telecom frequencies with a blackbody radiation shift $10^3$ times lower than current standards, erbium lattice clocks emerge as exciting candidates for modern metrology experiments\cite{kozlov_prospects_2013,patscheider_observation_2021}.

\begin{acknowledgments}

We thank Ana Maria Rey, Norman Yao, Philip Crowley, Maxwell Block, Bingtian Ye, and Nathan Leitao for insightful conversations about the structure of spin squeezing with finite-range interactions, squeezing with itinerance, and DTWA. We thank Maxwell Block for assistance benchmarking our simulations. We thank Yicheng Bao and Jim MacArthur for their advice on the microwave setup. We thank Jun Ye, Vladan Vuletic, and Peter Zoller for insightful conversations, and Brice Bakkali-Hassani, Lev Kendrik, Martin Lebrat, Philip Crowley, and Aaron Young for careful readings of this manuscript.

\textbf{Funding:}  We are supported by U.S. Department of Energy Quantum Systems Accelerator DE-AC02-05CH11231, National Science Foundation Center for Ultracold Atoms PHY-1734011, Army Research Office Defense University Research Instrumentation Program W911NF20-1-0104, Office of Naval Research Vannevar Bush Faculty Fellowship N00014-18-1-2863, Defense Advanced Research Projects Agency Optimization with Noisy Intermediate-Scale Quantum devices W911NF-20-1-0021, Army Research Office Grant W911NF20-1-0163, and Gordon and Betty Moore Foundation Grant GBMF11521. A.D. acknowledges support from the NSF Graduate Research Fellowship Program Grant DGE2140743. The computations in this paper were run in part on the FASRC Cannon cluster supported by the FAS Division of Science Research Computing Group at Harvard University.

\textbf{Author contributions:} A.D., V.K., L.S., M.S., and O.M. ran experiments and analyzed experimental data. A.D., V.K., L.S., M.S., V.S., and O.M. contributed to the experimental apparatus. A.D. and V.K. conceived the experiment. V.K. performed numerical simulations. M.G. supervised all works. All authors discussed the results and contributed to the manuscript.

\textbf{Competing interests:} M.G. is a cofounder and shareholder of QuEra Computing. All other authors declare no competing interests.

\textbf{Data availability:} The experimental data are available from the corresponding authors upon reasonable request.

\end{acknowledgments}

\bibliography{SpinSqueezing}

\newpage

\section*{Materials and Methods}

\setcounter{subsection}{0}

\subsection{Experimental Procedure}

\subsubsection{Band Insulator Preparation}

We start by preparing a degenerate Fermi gas (DFG) of $^{167}$Er, following the same two-stage MOT and forced evaporation sequence as our previous work \cite{phelps_sub-second_2020}. Our narrow line MOT and dipolar collision-dominated evaporation naturally leads to a spin-polarized 3D bulk gas in the state $\ket{g}=\ket{F=19/2,m_F=-19/2}$. The gas is then loaded into a \qty{532}{nm} accordion lattice and compressed, similar to our previous work with bosonic $^{168}$Er \cite{su_dipolar_2023}. Reducing the accordion power after compression induces evaporation in the vertical direction, removing any atoms in higher bands of the vertical lattice. Transferring from the accordion to the \qty{532}{nm} spacing vertical lattice is significantly more difficult than with the bosonic species of Er, requiring much better alignment between the accordion and vertical lattice, and twice the trap frequency, since the DFG is comparably incompressible. 

We then further evaporate the gas in 2D by "spilling over" in the x-y plane using a magnetic field gradient. We tune the atom number by adjusting the confining dimple size and power during spillover. We find that a spillover time of \qty{2.5}{s} leads to the coldest temperatures, roughly 10 times longer than the bosonic isotope $^{168}$Er, despite the elastic cross section only being $10\%$ smaller \cite{aikawa_reaching_2014, patscheider_determination_2022}. This is likely due to Pauli blocking, but may in part be due to a higher initial temperature. The magnetic field during spillover is quite sensitive due to the high density of Feschbach resonances \cite{frisch_quantum_2014}. We choose to operate at \qty{.55}{G}, which has a resonance-free range of $\pm$\qty{.05}{G}. 

We enforce arbitrary geometries of the atoms for Fig. \ref{fig:squeeze} by ramping up the DMD and ramping down the dimple while the system is still a bulk gas to minimize heating from transport. We then prepare a band insulator by adiabatically ramping up the x-y lattices and ramping down the dimple in \qty{400}{ms}. Each of the above steps was optimized using M-Loop \cite{wigley_fast_2016}.

\subsubsection{Spin State Preparation}
\begin{figure}[h]
    \includegraphics[width=0.5\textwidth]{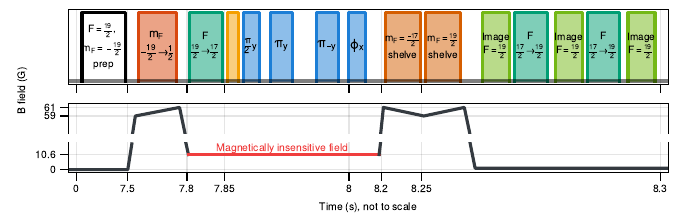}
     \caption{\textbf{Experimental control sequence.}
     Before spin manipulation, $^{167}$Er atoms are cooled spin-polarized into $\ket{F=19/2,m_F=-19/2}$. Following this preparation, the top row above shows the sequence of microwave pulses (with $\pi$ pulse pairs repeated every \qty{66}{ms}) in blue, microwave Landau-Zener sweeps in dark green, RF tones for Landau-Zener sweeps in red/orange, and imaging optical pulses in lime. The unlabelled yellow box is a blowout pulse that removes any atoms not in the target hyperfine states. The bottom row shows the magnetic field as a function of time. The interval during which the magnetic field is held at the magnetically insensitive point is highlighted in red. It is during this interval that the main squeezing procedure occurs. The magnetic field ramps to 60 G to enable $m_F\to m_F+1$ Landau-Zener sweeps with high fidelities, and the imaging is done at low field ($<1$ G).
     }
     \label{fig:expcontrol}
\end{figure}

The experimental procedure is shown in Fig. \ref{fig:expcontrol}. We transfer the spin states from  $\ket{g}=\ket{F=19/2,m_F=-19/2}$ to $\ket{\downarrow}=\ket{F=19/2,m_F=1/2}$ with a single Landau-Zener sweep through all 10 $m_F\to m_F+1$ transitions (red box), similarly to \cite{patscheider_controlling_2020}. To ensure high fidelity, we ramp the magnetic field to near 59 Gauss where the quadratic Zeeman effect induces a 90 kHz detuning between each adjacent $m_F\to m_F+1$ transition. We avoid Feschbach-induced losses from the $\sim26$ Feschbach resonances per Gauss by loading into a band insulator and ramping the magnetic field as fast as possible, and therefore diabatically with respect to tunneling \cite{frisch_quantum_2014}. Choosing a Rabi frequency near 2 kHz gives a total transfer fidelity of over 97\%. Furthermore, because our in-vacuum coil has a natural resonance at 60.8 MHz, we maintain the RF tone at that constant frequency while sweeping the magnetic field up by 2 Gauss over 100 ms to bring each of the 10 transitions in resonance with the RF tone.

Following the state preparation, we ramp the magnetic field to the magnetically insensitive value of \qty{10.622}{G} and apply a $\ket{\downarrow}\to\ket{\uparrow}=\ket{F=17/2,m_F=1/2}$ Landau-Zener sweep at 2.689 GHz (green box). The Rabi frequency of the sweep is 870 Hz, with a 20 ms duration and 30 kHz frequency ramp. The remaining atoms in $F=19/2$ (mostly in other $m_F$ states besides $m_F=1/2$) are then blown out with a resonant 583 nm frequency sweep across all $m_F$ states (yellow box).

Now that the atoms are spin-polarized in $\ket{\uparrow}$, the squeezing pulse sequence (blue boxes) can be applied. This typically consists of a $\pi/2$ pulse to prepare all atoms in the superposition $(\ket{\uparrow}+\ket{\downarrow})/\sqrt{2}$, followed by spin-echo $\pi$ pulses of alternating sign every 66 ms, ending with a readout pulse 90 degrees out of phase with the preparation pulse. 

\subsubsection{Spin Selective Imaging \label{sec:imaging}}

\begin{figure}[h]
    \includegraphics[width=0.5\textwidth]{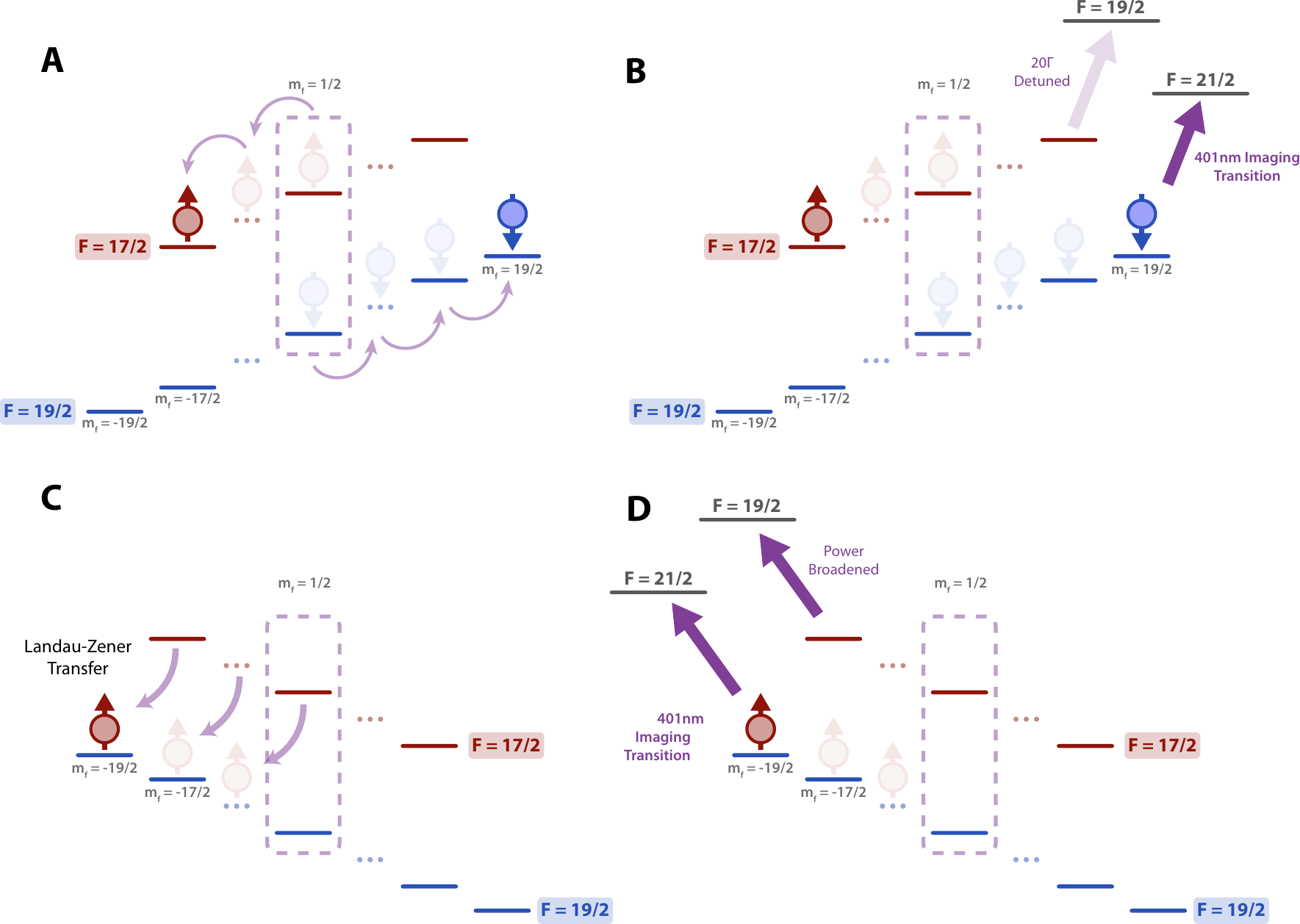}
     \caption{
     \textbf{Imaging sequence.}
     \textbf{a}, To prepare for imaging, the atoms in $F=17/2$ and $F=19/2$ are shelved in opposite stretched $m_F$ states.
     \textbf{b}, Atoms in $F=19/2$ are imaged first with a resonant optical pulse to $F=21/2$. This pulse is 20$\Gamma$ detuned from the corresponding transition for the atoms in $F=17/2$, and is further suppressed by the shelving due to Clebsch-Gordon coefficients.
     \textbf{c}, Atoms in all sublevels of $F=17/2$ are transferred to $F=19/2$.
     \textbf{d}, The atoms formerly in $F=17/2$ are imaged similarly to b, with higher power and shelving in the opposite stretched state.
     \label{fig:imaging}
     }
\end{figure}


After squeezing dynamics are complete we quench any itinerance by ramping the tunneling to \qty{.3}{Hz} in $\qty{100}{\mathrm{\mu s}}$. We then load atoms into a phase-matched \qty{488}{nm} accordion lattice. The lattice is expanded from the science lattice spacing of \qty{266}{nm} to an imaging spacing of \qty{2.5}{um} within \qty{60}{ms}, losing up to \qty{1}{\%} of the atoms in the process. We note that potential depth disorder, interactions, and tunneling during expansion conserve $\sigma_z$, and therefore our spin measurements later should remain unaffected. 

We then ramp the magnetic field up to \qty{60}{G} and perform Landau Zener sweeps to shelve the $\ket{\uparrow}$ atoms in the stretched state $\ket{F=17/2, m_F = -17/2}$, and the $\ket{\downarrow}$ atoms to the opposite stretched state $\ket{F = 19/2, m_F = 19/2}$ (Fig. \ref{fig:imaging} a). Subsequently, we ramp off the magnetic field and apply an optical pulse ($\sim .3$ $I_{sat}$) resonant with the \qty{30}{MHz} broad $F=19/2 \to F=21/2$ \; \qty{401}{nm} transition and \qty{20}{\Gamma} detuned from the respective $F=17/2 \to F=19/2$ transition to image all of the $\ket{\downarrow}$ atoms (Fig. \ref{fig:imaging}b). By placing the atoms in opposite stretched states, the Clebsh-Gordon coefficients shield the $F=17/2$ atoms from the optical pulse, suppressing potential off-resonant scattering by an additional factor of 50. We invert the in-plane magnetic field of our system to assign the negative $m_F$ stretched state more favorable Clebsh-Gordon coefficients (this corresponds to changing the handedness of the circularly polarized imaging light). Then we perform a Landau-Zener sweep to transfer the $F=17/2$ manifold to the $F=19/2$ manifold (Fig. \ref{fig:imaging}c) and once again apply a resonant optical pulse to image the atoms (Fig. \ref{fig:imaging}d). To ensure we imaged all atoms in our system, we reapply the same Landau-Zener transfer and then apply a power-broadened (20 $I_{sat}$) resonant optical pulse to image any atoms remaining in either of the two F manifolds. Typically we read $<.2$ \% of the atoms left for the final image.

We image on average 13 photons per atom across 144 pixels per atom to minimize the chance of exciting off-resonant transitions. We reduce dark counts and the time between images by operating our EMCCD camera (Andor iXon Ultra 897 EXF) at fast readout rates of \qty{17}{MHz}, resulting in dark counts equivalent to 2.3 photons per site. In addition, we minimize multiplication noise by setting the gain to maximum and binarizing each pixel, which works in the limit of much less than one photon per pixel \cite{su_fast_2024}. We obtain a $99.5$\% identification fidelity between 0 and 1 atom per site with the above parameters and less than $.25$\% atom number difference between all atoms in $\ket{\uparrow}$ and all atoms in $\ket{\downarrow}$ by fitting the Rabi oscillation in Fig. \ref{fig:rabi}, as well as the photon count overlap between the two well separated 0 and 1 atom peaks.

\subsection{Tunneling Methods}


We probe the effects of tunneling over three regimes: $t\approx\qty{0}{Hz},\qty{10}{Hz}$, and \qty{50}{Hz}. In the intermediate regime $t=\qty{10}{Hz}$, the tunneling rate into a single neighboring site is a factor of 4 greater than the collective interaction experienced by a single atom from the rest of the cloud. The tunneling is also much larger than the site-to-site disorder of \qty{3}{Hz} RMS. We turn off the DMD to minimize disorder, but that forbids us from using two clouds to subtract technical noise. To be confident in the measured squeezing values, we limit ourselves to clouds with less than 100 atoms. To directly compare the experimental results from different tunnelings, we keep the density the same by adiabatically ramping down the lattice from a band insulator, generating a relatively low-temperature state at $t=\qty{50}{Hz}$, and then quenching the lattice to the tunneling we want during squeezing. This keeps the fillings constant; however, for the $t=\qty{50}{Hz}$ data, because there is effectively no quench, the kinetic sector is substantially colder. Despite the assumption that increased tunneling would lead to more doubly occupied sites at finite polarization, and in turn inelastic collisions or dipolar relaxation, we experimentally measure that the rate of atom loss from the lattice is independent of tunneling by summing the two spin states in Fig. \ref{fig:tunneling}c. Even in the presence of collisional loss terms in the Hamiltonian, the high contrast measured, especially at $t=\qty{50}{Hz}$, indicates the cloud remains relatively spin-polarized during the dynamics, which should prevent collisions from occurring.

\subsection{Lattice Inhomogeneity in the Spin Sector}
For Fig. \ref{fig:mean} all data was taken with science lattice polarization aligned to roughly circular to remove any back reflections from the path using a beam splitter and quarter waveplate. The in-plane lattices generally had depths of $~35 E_R$ with corresponding $E_R$ of \qty{4.2}{kHz}, and the vertical lattice had a trap depth of $~31 E_R$ with an $E_R$ of \qty{1.05}{kHz}. The differential polarizability was measured to be $\qty{1.71}{Hz/E_R}$ and $\qty{2.36}{Hz/E_R}$ for the $x$ and $y$ science lattices in the above configuration, and $\qty{3}{Hz/E_R}$ for the vertical lattice. For squeezing experiments in Figures \ref{fig:squeeze}, \ref{fig:tunneling} we prioritized minimizing the differential polarizability, reducing it to less than $\qty{.1}{Hz/E_R}$ and $\qty{.2}{Hz/E_R}$ by rotating the $x$ and $y$ lattice polarizations respectively to tune the tensor and vector light shifts. We note that with retroreflected lattices it is impossible to create a magic polarization for the vertical lattice in the given dipole configuration. The differential polarizability leads to negligible site-to-site disorder, as we expect $\qty{\approx 30}{ppm}$ peak-to-peak disorder on the total potential, which is about $\qty{9}{mHz}$ in the spin sector, dominated by the vertical lattice. We also expect a global spin harmonic confinement dominated by the vertical lattice of approximately $\qty{1.7}{mHz/r^2}$ for Fig. \ref{fig:mean}, and $\qty{1.1}{mHz/r^2}$ after rotating to the magic polarization, where $r$ is the distance from the center of the potential in lattice sites, however, our spin echo sequences decouple the mean-field shift. Therefore the relevant parameter is the maximum site-to-site derivative, which is \qty{43}{mHz} for 1600 atoms; small enough to not affect dynamics according to our simulations, even for our largest systems. However, we still included the potential in the DTWA for Fig. \ref{fig:squeeze}.

\subsection{Magnetically Insensitive Transition}

The chosen qubit between $F=\ket{19/2,m_F=1/2}$ and $\ket{F=17/2,m_F=1/2}$ is first-order magnetically insensitive, with a second-order coefficient of $1.2\:\mathrm{mHz/mG}^2$ at an out-of-plane quantization field of \qty{10.622}{G} (Fig. \ref{fig:mainexp}c). 


Magnetic insensitivity prohibits effective Ising interactions, as magnetically insensitive atoms are not affected by the mean local B field generated by the existence of another atom, but the chosen $\ket{F,m_F}$ qubit states rotate the native Ising interaction between different $m_J$ projections into spin-exchange interactions in the hyperfine manifold due to $^{167}$Er's nonzero nuclear spin \cite{zou_magnetic_2020} (see Section \ref{sec:origin}).

Because of the constraint of magnetic insensitivity, the strength of the interaction is significantly reduced. For $^{167}$Er, the maximum spin-exchange magnetic interaction on our lattice is \qty{9.3}{Hz}, between $\ket{F=19/2,m_F=1/2}$ and $\ket{F=19/2,m_F=-1/2}$, while the family of magnetically insensitive transitions has interactions ranging from \qty{1.09}{\Hz} to \qty{3.7}{\Hz} (Fig. \ref{fig:dipolarstrength}). We measure our magnetic field noise to have a RMS on the order of 1 mG, which limits the coherence time of a qubit between $\ket{F=19/2,m_F=1/2}$ and $\ket{F=19/2,m_F=-1/2}$ to 1 ms. This would result in an interaction strength with a coherence time product 700 times lower than what we measure with the magnetically insensitive qubit.

We choose the specific qubit with an interaction strength of \qty{1.09}{Hz} because of the ease of state preparation and readout with those states; our atoms are natively cooled and imaged in $F=19/2$. Note that should we have operated with this qubit at \qty{0}{G}, the magnetic sensitivity would be $25\: \mathrm{kHz/G}$, leading to a coherence time of 40 ms, a reduction by over 2 orders of magnitude from what we measure.

\subsection{$T_2$ and Lattice Loss Measurements}

\begin{figure}[h]
    \includegraphics[width=0.35\textwidth]{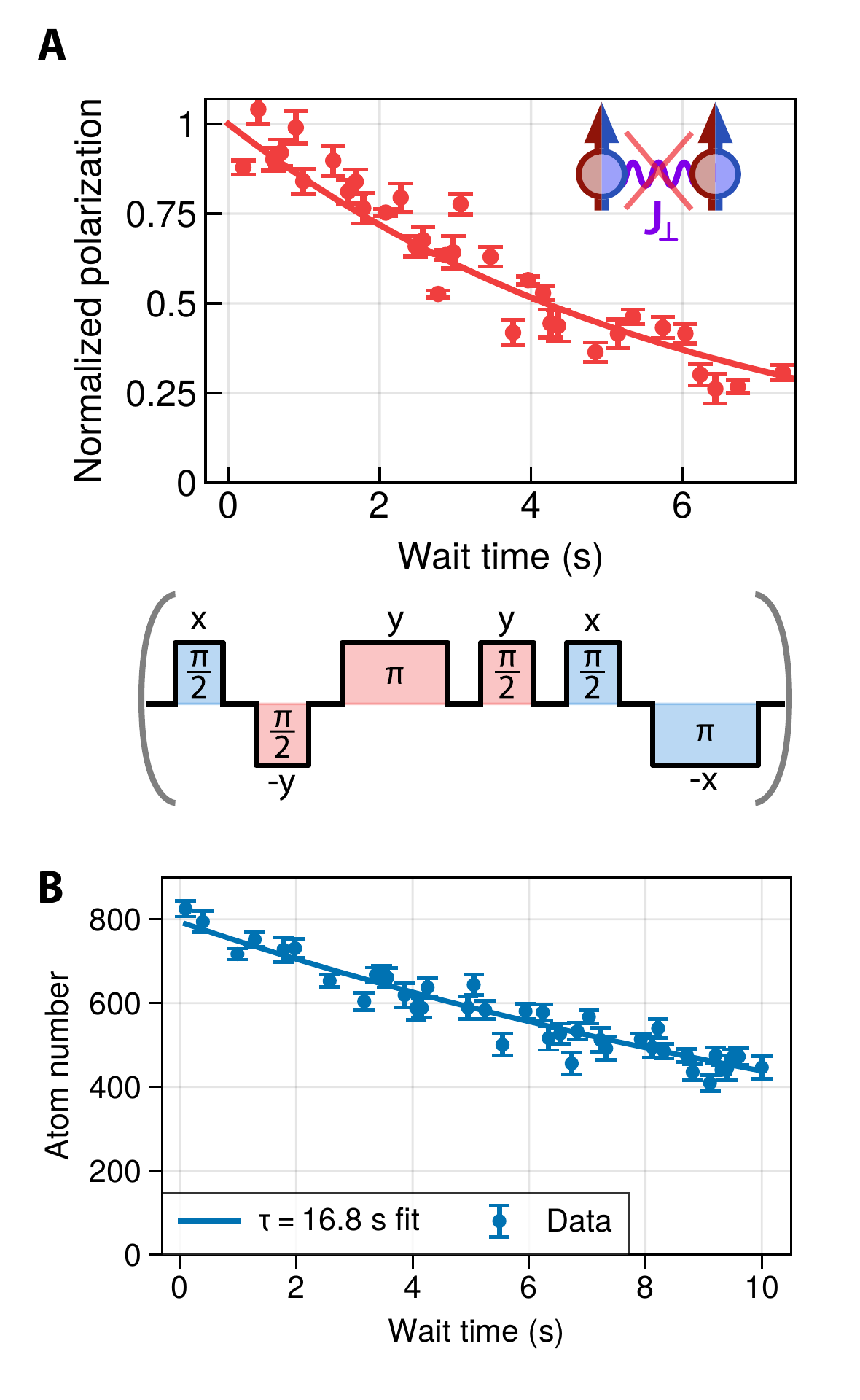}
     \caption{
     \textbf{Lattice lifetime and $T_2$ Coherence }
     \textbf{a}, $T_2$ Coherence of a cloud of 900 atoms with disorder and interaction effects decoupled with a spin echo + WAHUHA pulse sequence. We measure a $T_2$ of $\qty{6.0(2)}{s}$. The pulse sequence is shown below the data; the sequence within the grey parentheses is repeated every \qty{66}{ms}. \textbf{b} We measure the single-particle lattice loss and fit to an exponential to obtain a lattice lifetime of 16.8 s. 
     }
     \label{fig:latticeloss}
\end{figure}

In Fig. \ref{fig:latticeloss}a, we show a $T_2$ measurement without spin resolved imaging by computing a lower bound on the state polarization, normalized for atom loss from our lattice. We perform this measurement by first measuring the atom loss and fitting an exponential (Fig. \ref{fig:latticeloss}b), and then measuring the contrast $S_z/|S|$ of the same cloud after dynamical decoupling from the spin echo + WAHUHA pulse sequence. We divide the latter by the fit of the former to obtain the normalized polarization, to which we fit an exponential and extract the $T_2$ time. We observe loss rates from the lattice that are independent of the initial spin state, and that are unchanged after dephasing the atoms by ramping a magnetic field gradient or applying a large optical potential gradient.

\subsection{Data analysis}

From each experimental shot, we extract spin-dependent lattice occupation matrices using our site-resolved imaging as described in Section \ref{sec:imaging}. From this, we compute occupation numbers $N_{\uparrow}, N_{\downarrow}$ for each shot. For each group of scans that are analyzed together, we discard all shots whose atom numbers are not between the 12.5 and 87.5 percentile of atom numbers in that group, and across all shots, we do not factor in lattice sites whose mean occupation is less than 10\% of the maximum mean occupation within that scan group. Atom number fluctuations greater than one std. are dominated by long-term drift. We analyze each observable as described below.

\subsubsection{Contrast}

The Ramsey contrast $C$ is measured by preparing the state on the equator of the generalized Bloch sphere and applying a $\pi/2$ readout pulse with varying phase $\phi$. We then fit a $\sin$ function to the resulting ratio $S_z/S= (N_{\uparrow}-N_{\downarrow})/N$ as a function of $\phi$, with the data uncertainty being the standard error of the mean. The amplitude of this fit is then $C$. To account for data where $C$ is close to 1 (and thus when the underlying distribution of $S_z/S$ is far from Normal) we bootstrap the data (500 resamples) and perform the fit for each resample, after which we obtain a distribution for $C$ which we clip to the range $[0,1]$. We report the median of this distribution as our estimate for $C$ rather than the fit of the original data to account for the fact that this fitting procedure can be sensitive to small perturbations in the data. Error bars are then computed from the 16\% and 84\% percentiles. In the case when $S_z/S$ is exactly equal to 1 for all shots at a particular readout angle, we set the s.e.m. of that point to $0.01$ for fit convergence.

\subsubsection{Spin Squeezing}

For direct squeezing measurements (Fig. \ref{fig:tunneling}) we estimate the spin projection noise $\xi^2$ by directly sampling $m$ shots. From each shot we extract the spin populations $N_{\uparrow}$ and $N_{\downarrow}$. We then estimate $\xi^2$ with
\begin{equation}
    \frac{\text{Var}[N_{\uparrow}-N_{\downarrow}]}{\hat{\sigma}^2_{\mathrm{SQL}}}
\end{equation}
where the estimate of the SQL, $\hat{\sigma}^2_{\mathrm{SQL}}$, is defined as:
\begin{equation}
    \begin{aligned}
        \hat{\sigma}^2_{\mathrm{SQL}}\equiv 4\langle p\rangle(1-\langle p\rangle)\langle N\rangle,\\
        \langle p\rangle\equiv(1/2)(1+\langle(N_{\uparrow}-N_{\downarrow})/N\rangle)
    \end{aligned}
\end{equation}
and $\langle\cdot\rangle$ denotes an average over the sample. We use this definition to account for potential systematic pulse errors that might result in $\langle N_{\uparrow}-N_{\downarrow}\rangle\neq 0$ at readout. The variance term is computed as a simple sample variance. To obtain error bars on our estimator, we bootstrap the sample (1000 resamples) and calculate the 16\% and 84\% percentiles.

For differential squeezing measurements (Fig. \ref{fig:squeeze}) we prepare two spatially separated clouds $A, B$ and extract spin population numbers from each by manually defining a separation line. $\xi^2$ is then computed as:
\begin{equation}
   \xi^2\equiv
    \frac{\text{Var}[(N^{(A)}_{\uparrow}-N^{(A)}_{\downarrow}) - (N^{(B)}_{\uparrow}-N^{(B)}_{\downarrow})]}{\hat{\sigma}^{2,(A)}_{\mathrm{SQL}}+\hat{\sigma}^{2,(B)}_{\mathrm{SQL}}}
\end{equation}
where $\hat{\sigma}^{2,(A,B)}_{\mathrm{SQL}}$ is the same as in the non-differential case, but for each of the two clouds.

To compute error bars on the interferometric squeezing $\xi^2_R$ we combine bootstrapped distributions of $\xi^2$ and the contrast to obtain a bootstrapped distribution (500 resamples) of $\xi^2_R$, from which we compute the 16\% and 84\% percentiles.

To compute the minimum squeezing from a dataset of $\xi^2$ or $\xi^2_R$ as a function of readout angle, we curve fit to either a $\sin$ function (Fig. \ref{fig:squeeze}), or a quadratic (if the range of the sampled $\theta$ points does not extend significantly beyond the approximately quadratic region of the $\sin$, as in Fig. \ref{fig:tunneling}). We note that any state, when rotated around the Bloch sphere, will always exhibit an elliptic variance, and therefore the squeezing can be fitted to a squared $\sin$ with an offset. Antisqueezed points determine the phase of the $\sin$ function, and measured squeezed points are then interpolated to determine the offset and thus variance that would experimentally be measured. The primary source of deviation from a sinusoidal model would be any systematics as a function of rotation angle in the final rotation pulse, but we measure our pulses to be sufficiently good to ignore this effect from both Rabi oscillations and $\pi$ pulse fidelities. 

The uncertainty of the points for the fits are set to half the difference between the 84\% and 16\% percentiles of the bootstrapped distributions of $\xi^2$ or $\xi^2_R$. To obtain the upper and lower fit error lines, we sample 1000 curves from the covariance matrix of the fitted parameters and plot the 84\% and 16\% percentiles for every readout angle. The final reported squeezing is extracted from the fit parameters, with error bars from the fit covariance matrix. This errorbar is then propagated in the standard manner into dB.

For differential measurements, as a function of evolution time the optimally squeezed readout angle approaches zero, but the XY symmetric interaction, for an isolated cloud, preserves the original variance along $S_z$, which means that the optimal squeezing angle should never reach zero. However, our theory and experiment both show that for $\tau>\qty{500}{ms}$ the optimal squeezing angle occurs near zero. We confirm this is an artifact of the finite dipolar coupling between the two clouds by observing that the effect disappears upon removing one cloud from the numerics. 

\subsubsection{Correlations}

We compute site-to-site two particle correlations $g^{2}_{i,j}\equiv \langle \hat{\sigma}^z_i\hat{\sigma}^z_j\rangle-\langle \hat{\sigma}^z_i \rangle \langle\hat{\sigma}^z_j\rangle$, where $\langle \hat{\sigma}^z_i\rangle$ is an average over all shots with atoms in site $i$, and $\langle \hat{\sigma}^z_i\hat{\sigma}^z_j\rangle$ is an average over all shots with atoms in sites $i$ and $j$ (other shots are discarded). Then for each $i$, we select all $j$'s within a 9x9 site square centered around $i$, and we average over every such subsection of sites within the cloud. The resulting $9\times 9$ plot is symmetrized by averaging over all four $90$ degree rotations, followed by averaging with itself flipped over its diagonal. We then show a 9x9 2D plot using a stitched linear-log scale to ensure that the structure in the correlations is still visible even if the absolute magnitudes of the correlations are small.

\subsection{Comparisons to theory (DTWA)}

Throughout this paper, we compare our results to numerics implemented with the Discrete Truncated Wigner Approximation (DTWA) \cite{schachenmayer_many-body_2015}. This semi-classical method captures the buildup of quantum correlations through the evolution of an entangling Hamiltonian from an initially uncorrelated, classical state. For each experimental dataset, we randomly sample 10 measured fillings as the initial state. For each of these 10 fillings, we compute 100 classical trajectories, sampling the initial uncorrelated state. From these, we extract the relevant spin observables, such as contrast and squeezing. Error bars, shown in the thickness of the shaded theory curves, are calculated as standard errors of the mean over the 10 shots. When comparing metrological squeezing $\xi^2_R$ results to DTWA, we compute $\xi^2_R$ from the DTWA simulations of the squeezing dataset (rather than the contrast dataset).

Due to the finite width of the atomic Wannier functions in the $z$ axis compared to the xy lattice spacing, the atoms do not experience the full dipolar interaction at short distances. We account for the reduction in the mean squeezing field by rescaling the total interaction to the predicted 86\% (90.6\% in Fig. \ref{fig:mean}a) but do not incorporate the change in spatial dependence of the potential. In the DTWA simulations of squeezing without tunneling (Figure \ref{fig:squeeze}), we incorporate the $\qty{1}{mHz/site^2}$ harmonic disorder, and the spin-echo $\pi$ pulses every 66 ms that decouple this disorder.

For the intermediate tunneling regime ($t=10$ Hz), we incorporate diffusive tunneling in DTWA using stochastic hopping based on the Metropolis-Hastings algorithm. While we expect coherent tunneling in the experiment, we assume that we start from a high temperature occupation of momentum modes because of the lattice depth quench, and therefore experience effectively classical motion. We consider only nearest-neighbor tunneling at a rate $t$ (in Hz), in a DTWA simulation with step size $dt$ (in seconds). Each time step, we propose for an atom in lattice site $i$ to hop to a random empty adjacent site along the $x$ axis with probability $4\:m\:t\:dt$ where $m$ is the number of empty nearest-neighbor sites along the $x$ axis (at most 2). The factor of 4 comes from the rate of ballistic expansion of a free particle in a lattice, where $1/(2t)$ is the time to do a full oscillation for a two site system \cite{preiss_strongly_2015}. We then accept this proposal with probability $\mathrm{min}\{1, P(j)/P(i)\}$, where $P$ is the experimentally observed averaged lattice filling of the dataset being simulated to account for harmonic confinement and disorder. This procedure is repeated for the $y$ axis within the same timestep. The acceptance factor ensures the lattice filling of the simulation approximates the filling observed experimentally even while the atoms undergo hopping. To smooth discontinuities in the acceptance function, we apply a one site std. Gaussian filter over the experimental filling distribution to obtain $P$. We benchmark our algorithm by ensuring that the final simulated lattice filling matches that observed experimentally at the end of the simulation. 

For the high tunneling regime ($t=50$ Hz), we simulate an all-to-all OAT model with DTWA by replacing the strength of the XX+YY interaction between every pair of atoms with the average value of this strength over the experimentally observed cloud. Further theoretical investigations can be modified to include quantum fluctuations and effective temperature in the motion using the fermionic truncated Wigner Approximation
\cite{davidson_semiclassical_2017}.

\subsection{Microwave and RF setup}

\begin{figure}[h]
    \includegraphics[width=0.5\textwidth]{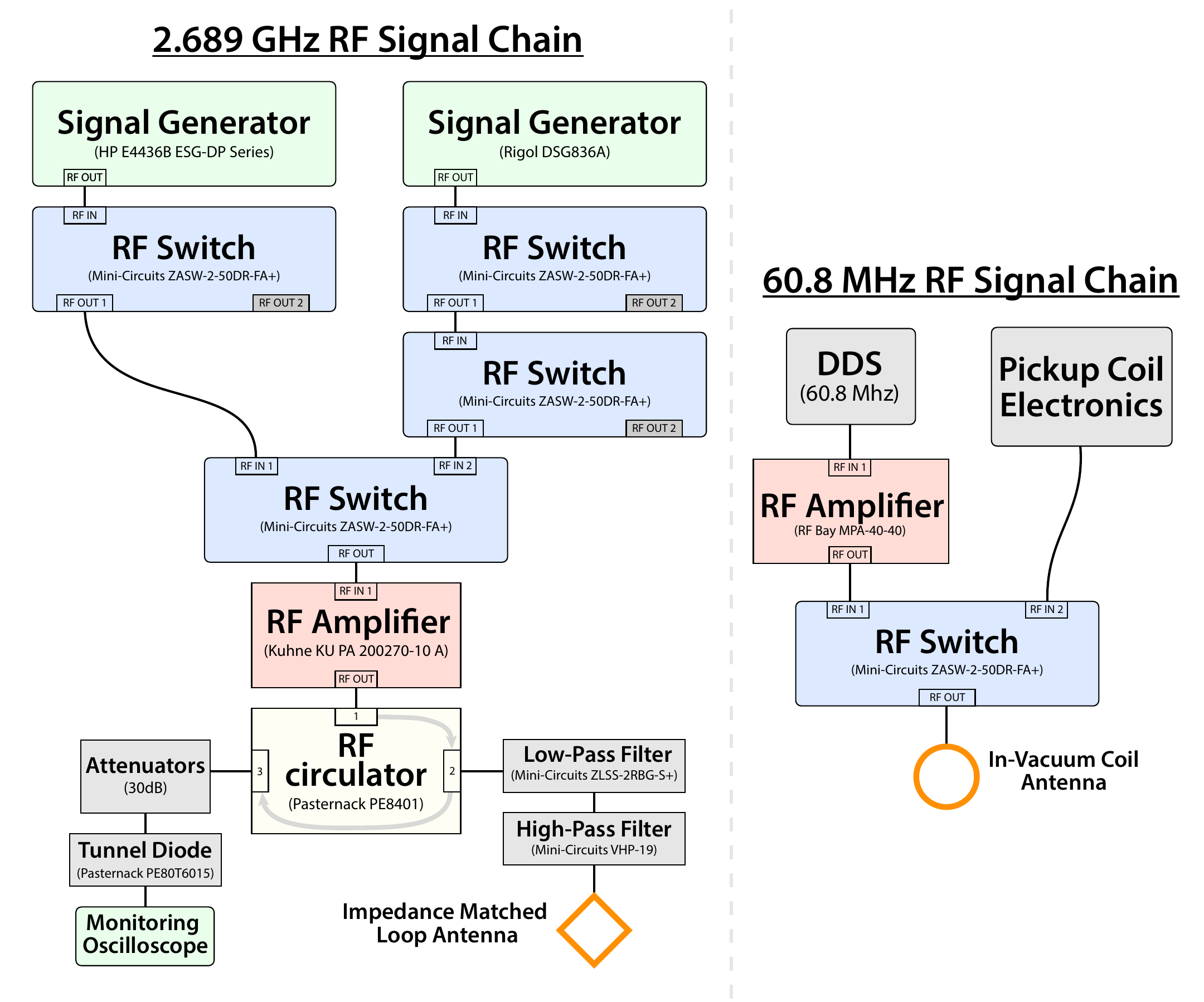}
     \caption{
     \textbf{RF signal chain.}
    The RF components that deliver microwave and RF tones through two antennae into the science chamber. A loop antenna (gold square) is used for the main qubit frequency of 2.689 GHz, while a pair of large coils (gold circle) in vacuum transmit 60.8 MHz tones, and pick up the 2.689 GHz tones emitted by the former antenna for monitoring.
     }
     \label{fig:rf}
\end{figure}

In order to precisely manipulate the hyperfine states of our atoms, we generate a variety of microwave pulses and microwave/RF frequency sweeps. We detail the components of this setup, with part numbers, in Fig. \ref{fig:rf}. 

For transfer between $m_F$ levels, we generate a 60.8 MHz RF signal to drive an in-vacuum magnetic field coil that acts as an antenna. This 60.8 MHz is generated using an in-house built DDS (Direct Digital Synthesizer) which is amplified and gated with an RF switch.

\begin{figure}[h]
    \includegraphics[width=0.5\textwidth]{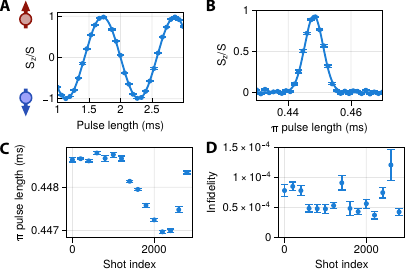}
     \caption{
     \textbf{Coherent microwave spin control.}
     \textbf{a}, 870 Hz Rabi oscillations of the hyperfine qubit, fitted to a $\sin$ with an amplitude of $.9882(7)$, and an offset of $.0051(6)$.
     \textbf{b}, Spin polarization after 400 consecutive $\pi$ pulses as a function of pulse length, fitted to a Gaussian with a linewidth of $\qty{\approx 3}{\mu s}$. 
     \textbf{c}, Drift in the Rabi frequency overnight.
     \textbf{d}, Drift in the $\pi$ pulse infidelity overnight.
     }
     \label{fig:rabi}
\end{figure}

To manipulate our atoms within the spin squeezing manifold we generate a 2.689 GHz RF signal corresponding to the splitting between the states of the hyperfine qubit. This signal couples directly to the qubit magnetic dipole moment with a Rabi frequency of \qty{870}{Hz} (Fig. \ref{fig:rabi}a) and a $\pi$ pulse fidelity lower bound of \qty{.999981 (3)}{}, measured through 2000 repeated same phase $\pi$ pulses with \qty{2}{ms} waits in between over \qty{5}{s}. For this fidelity estimate, the measured feature is the amplitude of a Gaussian centered at the optimal $\pi$ pulse duration, with a standard deviation of \qty{3}{us}, similar to Fig. \ref{fig:rabi}b. The effective Gaussian occurs from phase fluctuations and the XY interaction dephasing the atoms unless the atoms spend most of their time at the poles of the generalized Bloch sphere. 

By repeated measurements of 400 repeated pulses with \qty{4}{ms} waits, we follow the peak of the Gaussian in batches of 200 shots over 3000 overnight shots. We measure an overnight drift of $.4\%$ (Fig. \ref{fig:rabi} c), which limits us to a long-term fidelity of \qty{.99984 (1)}{} (averaged over the whole scan) without feedbacking the Rabi frequency (note the instantaneous fidelity is lower than the quoted $\approx 10^{-4}$ due to the binning of more shots), shown in Fig. \ref{fig:rabi} d.

We conduct both discrete microwave rotations and Landau-Zener transfers between the two qubit states for state preparation, spin squeezing, and readout. We have two signal generators dedicated to the qubit manipulation. For conducting discrete microwave rotations we derive our signal from a stable source at a constant frequency. For Landau-Zener transfers we use a separate arbitrary waveform generator that allows for easy programming of frequency sweeps. We switch between these two signals using an RF Switch and subsequently pass them through a low-noise amplifier. These signals are fed into a loop antenna mounted to the exterior of our vacuum chamber. This antenna is impedance-matched with a shunt to provide $>99$\% transmission at the target frequency. 

Additionally, to prevent any unwanted precession or state transfer in the qubit we require a significant extinction ratio between our signal generators and our antennae when no signal is desired. This is accomplished by chaining together RF switches to gate the signals and execute sharp and reliable signal cut-off. Furthermore, the \qty{2.689}{GHz} qubit signal is bound by stringent phase noise requirements. Phase jitter on the order of 1 deg would result in a maximal viable noise squeezing measurement on the order of \qty{-10}{dB} for 500 atoms without a differential measurement. To combat this our signal generators for the qubit manipulation signal are locked to an ultra-low phase noise 10 MHz oscillator (Wenzel 501-31686). 

To test and monitor our main out-of-vacuum antenna we have two monitoring setups. A circulator is installed to track the back-reflection from the antenna with a tunnel diode, while the in-vacuum magnetic field coil is also used to measure the transmission of the 2.689 GHz signal into the vacuum chamber when it is not being used to perform frequency sweeps.

\subsection{Interspin Onsite Interaction}

We attempted to measure the on-site interspin interaction by preparing a superposition of $\ket{\uparrow}$ and $\ket{\downarrow}$ at a bias field of ~\qty{1}{Gauss}, far away from the magnetically insensitive point. We apply a magnetic field gradient for \qty{600}{ms} to dephase the spins since two spin-polarized fermions cannot occupy the same site. After dephasing the spins we modulate the amplitude of one of the science lattices for \qty{2}{s} in an attempt to drive a transition from one particle on two sites to two particles on one site, resulting in heating. However, we do not find any resonances at amplitude modulation frequencies between \qty{0}{kHz} and \qty{32}{kHz} that do not disappear in a spin-pure sample. We find it unlikely that the on-site interaction would be so large but leave a more careful study for future work. We note that our inability to measure the on-site interaction likely implies that it has little bearing on the dynamics of the paper.

\subsection{Origin of the Spin-Exchange Interaction \label{sec:origin}}

\begin{figure}[h]
    \includegraphics[width=0.5\textwidth]{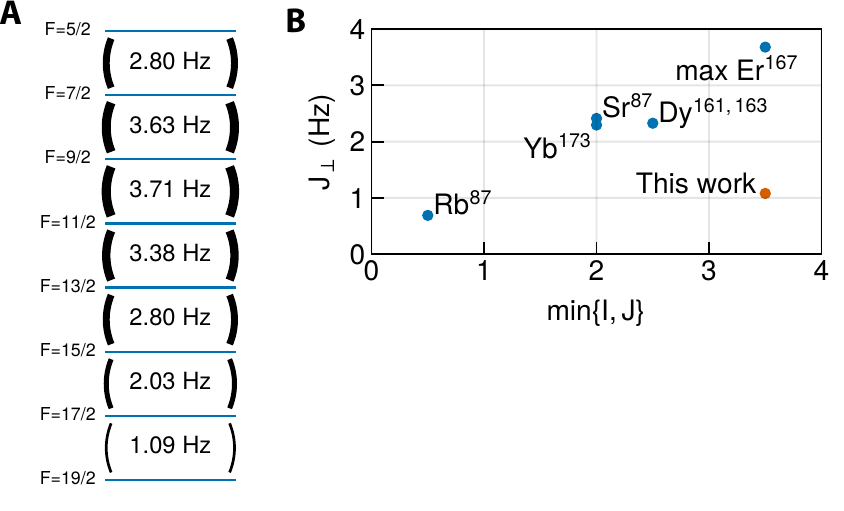}
     \caption{
     \textbf{Magnetically insensitive dipolar interaction strength in various elements.}
     \textbf{a}, Maximum interaction strength between every adjacent pair of $F,F+1$ manifolds in $^{167}$Er, which is always achieved in the $m_F=\pm1/2$ sublevels.
     \textbf{b}, Maximum nearest-neighbor interaction $J_{\perp}$ of the form used in this work in various fermionic isotopes (and $^{87}$Rb) relevant to cold atom experiments, assuming a 266 nm lattice spacing. The interaction increases with $\min\{I,J\}$, where $I,J$ are the nuclear and electronic spin quantum numbers respectively.
     }
     \label{fig:dipolarstrength}
\end{figure}

The interaction Hamiltonian between two magnetic dipoles $\mu_1,\mu_2$ a distance $r$ in direction $\hat{r}$ away from each other is
\begin{eqnarray}
    H_{dd}=-\frac{\mu_0}{4\pi r^3}[3(\vec{\mu}_1\cdot\hat{r})(\vec{\mu}_2\cdot\hat{r})-\vec{\mu}_1\cdot\vec{\mu}_2]
    \label{eq:dd}
\end{eqnarray}
For magnetic atoms, the dipole moment operator can be written as
\begin{eqnarray}
  \vec{\mu}=-\frac{\mu_B}{\hbar}(g_S\vec{S} + g_L\vec{L})-\frac{\mu_N}{\hbar}g_I\vec{I}
\end{eqnarray}
where $\mu_B,\mu_N$ are the Bohr and nuclear magnetons, respectively, and $\vec{S},\vec{L},\vec{I}$ are the spin, orbital, and nuclear angular momenta, respectively.
At all experimentally relevant fields, the $\vec{L}\cdot\vec{S}$ coupling in Erbium dominates and the $\vec{L}$ and $\vec{S}$ contributions to $\vec{\mu}$ can be viewed through a single $g_J\vec{J}$ factor. In the ground state of Erbium, $g_J\approx 1.16$ \cite{frisch_dipolar_2014}. Furthermore, since the ratio of the Bohr magneton to the nuclear magneton is on the order of 2,000, we can safely neglect $\vec{I}$ in calculations of magnetic coupling between atoms. Thus we can write the magnetic moment approximately as
\begin{eqnarray}
    \vec{\mu}\approx -\frac{\mu_Bg_J}{\hbar}\vec{J}
\end{eqnarray}
and thus the dipole-dipole Hamiltonian can be written approximately as
\begin{eqnarray}
    H_{dd}\approx -\frac{\mu_0 g_J^2\mu_B^2}{4\pi r^3\hbar^2}[3(\vec{J}_1\cdot\hat{r})(\vec{J}_2\cdot\hat{r})-\vec{J}_1\cdot\vec{J}_2]
    \label{eq:dd2}
\end{eqnarray}
In the experiment, we arrange $\text{Er}^{167}$ in a 2D lattice along $\hat{x},\hat{y}$ with spacing $a=$ \qty{266}{nm}. We apply an out of plane quantization field $\vec{B}=B\hat{z}$. For small magnetic fields ($<100$ G), the $\vec{I}\cdot\vec{J}$ hyperfine coupling dominates the Zeeman term $-\vec{B}\cdot\vec{\mu}$ in the single-atom Hamiltonian. Thus the single-atom eigenstates are approximately $F$-states. We define a qubit between two states with $\Delta F=1$ but $\Delta m_F=0$, which allows us to find a ``magic" magnetic field at which we are first-order magnetically insensitive. We consider the interaction of (\ref{eq:dd2}) in this qubit manifold. Since $\Delta m_F=0$, these states are not coupled by any terms that alter $m_J$ without altering $m_I$:
\begin{eqnarray}
    \bra{F+1,m_F}\vec{J}_{\pm}\ket{F,m_F}=0
\end{eqnarray}
Since $J_x,J_y$ are linear combinations of $J_{\pm}$ the only term with non-vanishing coupling contribution in (\ref{eq:dd2}) is $J_{1z} J_{2z}$. This yields a nearest-neighbor spin exchange coupling of
\begin{eqnarray}
    C\equiv \bra{F+1}_1\bra{F}_2\frac{\mu_0 g_J^2\mu_B^2}{4\pi a^3\hbar^2}J_{1z}J_{2z} \ket{F}_1\ket{F+1}_2\\
    =\frac{\mu_0 g_J^2\mu_B^2}{4\pi a^3}\left|\bra{F+1,m_F}\frac{J_z}{\hbar}\ket{F,m_F}\right|^2
\end{eqnarray}
In the experiment,
\begin{eqnarray}
    \frac{1}{\hbar}\frac{\mu_0 g_J^2\mu_B^2}{4\pi a^3}\approx (2\pi\times)\:0.934\:\text{Hz}
\end{eqnarray}
while the term
\begin{eqnarray}
    C_G\equiv\left|\bra{F+1,m_F}\frac{J_z}{\hbar}\ket{F,m_F}\right|^2=\qquad\\
    \left(\sum_{m_J=-J}^{J}m_J\mathcal{C}^{J,I,F}_{m_J,(m_F-m_J),m_F}\mathcal{C}^{J,I,(F+1)}_{m_J,(m_F-m_J),m_F}\right)^2
\end{eqnarray}
depends on the Clebsch-Gordon coefficients $\mathcal{C}_{m_1,m_2,m}^{j_1,j_2,j}$ associated with the $F,m_F$ used. Our experimental sequence utilizes $F=17/2,F+1=19/2,m_F=1/2$. For these states, $C_G=\frac{420}{361}$, giving a nearest-neighbor exchange term of $1.087\;\text{Hz}$. At the magic B field, there is no single particle $Z$ expectation value, and therefore there cannot be a $ZZ$ Ising term. Note that the above derivation was done at zero field, not the magic B field, so we expect the interaction to be slightly different due to the eigenstates no longer being pure $F,m_F$ states. We plot the available interactions of this form in $^{167}$Er, as well as several other atomic isotopes relevant to cold atom experiments, in Fig. \ref{fig:dipolarstrength}.

\end{document}